# Quantum Interference between Autonomous Single-Photon Sources from Doppler-Broadened Atomic Ensemble

Taek Jeong, Yoon-Seok Lee, Jiho Park, Heonoh Kim, and Han Seb Moon*

*Department of Physics, Pusan National University, Geumjeong-Gu, Busan 46241, South Korea*
*Corresponding author: hsmoon@pusan.ac.kr*

**To realize a quantum network based on quantum entanglement swapping, bright and completely autonomous sources are essentially required. Here, we experimentally demonstrate Hong-Ou-Mandel (HOM) interference [1] between two independent bright photon-pair sources generated via the spontaneous four-wave mixing in Doppler-broadened ladder-type $^{87}$Rb atoms. Bright autonomous heralded single-photon sources are operated in a continuous-wave (CW) mode with no synchronization or supplemental filters. The four-fold photon coincidence counts per hour correspond to 200 events for a temporal range of 0.4 ns, higher than those of previously reported for autonomous single photons [2, 3]. We observe an HOM dip with 83% visibility with the two autonomous heralded single-photons for an effective measurement time of 900 s. The achievement of HOM interference between the two bright autonomous photon-pair sources constitutes an important step towards a practical scalable quantum network.**

Since the first proposal of entanglement swapping [4], quantum interference between independent single photons has played an important role in both long-distance quantum communication incorporating quantum repeaters and optical quantum-information processing by a linear optics quantum computer, towards realization of a scalable quantum network [5–8]. In particular, Hong-Ou-Mandel (HOM) quantum interference constitutes key evidence of the indistinguishability and purity of independent single-photon sources [9]. Such indistinguishable independent single photons must have identical spectral, spatial, polarization, and temporal modes at a beam splitter [10].

Spatial overlap can be achieved through the use of single-mode optical fibers, and polarization can be matched by a polarization controller. For temporal mode matching, previous experiments involving independent single-photon sources have used heralded single-photon sources operated in the pulse mode [10–14]. In these scenarios, synchronized pulse sources created two photons simultaneously, and path length matching was employed to obtain the required temporal overlap. However, the pulsed heralded single photons produced in this manner cannot be entirely autonomous, because of the indispensable synchronization [2, 3].

In addition, spectral overlap and spectral narrowing between two independent single-photon sources is necessary for the observation of HOM interference. In the case of two independent heralded single photon sources produced via spontaneous parametric down-conversion (SPDC), narrow bandpass filters have been used to achieve spectral overlap and spectral narrowing [2, 10–14]. For narrow-band single-photon sources generated from atomic ensembles [3, 15–21], however, no bandpass filter is required to achieve HOM interference, because the coherence time of the single-photon source is longer than the time resolution of a single-photon detector (SPD). In particular, the pure temporal state of the single-photon source is as important as the time-frequency entanglement source [3].

Another important consideration is the brightness of the heralded single photon sources. In previous experiments [2, 3], a low four-fold photon coincidence count rate between two autonomous photon-pair sources was obtained, corresponding to 1 events/h for a nonlinear crystal [2] and 40 events/h for a cold atomic ensemble [3], respectively. In particular, achieving a bright autonomous photon-pair source without the use of a narrow bandpass filter for application in a practical quantum scalable network is challenging.

In this Letter, we experimentally demonstrate HOM interference between two independent continuous-wave (CW) photon-pair sources, which are bright and completely autonomous. The CW photon-pair sources, which are generated via the spontaneous four-wave mixing (SFWM) process of a Doppler-broadened atomic ensemble, allow the use of completely autonomous sources in a warm atomic ensemble for the first time. We characterize the bright autonomous photon-pair sources and demonstrate that HOM interference using two independent CW single-photon sources is realized with no synchronization or supplemental filters. In addition, the four-fold photon coincidence count rate is compared with those previously reported for autonomous single photon sources [2,3].

Figure 1(a) shows the ladder-type atomic configuration for photon-pair generation used in this study, which employs the $5S_{1/2}$–$5P_{3/2}$–$5D_{5/2}$ transition of $^{87}$Rb. To minimize the fluorescence due to the one-photon resonance, the pump laser is blue detuned by approximately 800 MHz from the $5S_{1/2}$ (F = 2)–$5P_{3/2}$ (F' = 3) transition. Then, the coupling field is red detuned by approximately 800 MHz from the $5P_{3/2}$ (F' = 3)–$5D_{5/2}$ (F" = 4) transition, so as to satisfy the two-photon resonance of the $5S_{1/2}$ (F = 2)–$5P_{3/2}$ (F' = 3)–$5D_{5/2}$ (F" = 4) transition. Owing to two-photon coherence on the two-photon resonance, the signal photon of the $5P_{3/2}$ (F' = 3)–$5D_{5/2}$ (F" = 4) transition and the idler photon of the $5S_{1/2}$ (F = 2)–$5P_{3/2}$ (F' = 3) transition are emitted from the Doppler-broadened ladder-type atomic system shown in Fig. 1(a), via the SFWM process. In particular, note that our photon-pair source can be operated in the CW mode, because no preparation or optical pumping processes are required to obtain high correlation between the photon pairs. The main cause of the high brightness is the collective enhancement of the emission rate due to the coherent contributions from almost all the velocity classes in the Doppler-broadened atomic ensemble [22].

cps/mW. Further, the signal- and idler-photon single counts and the photon-pair coincidence counts increased linearly with increasing pump power. The non-classical correlation between the photon pairs indicated strong violation of the Cauchy-Schwarz inequality by a factor $R$ as high as 2000 [22].

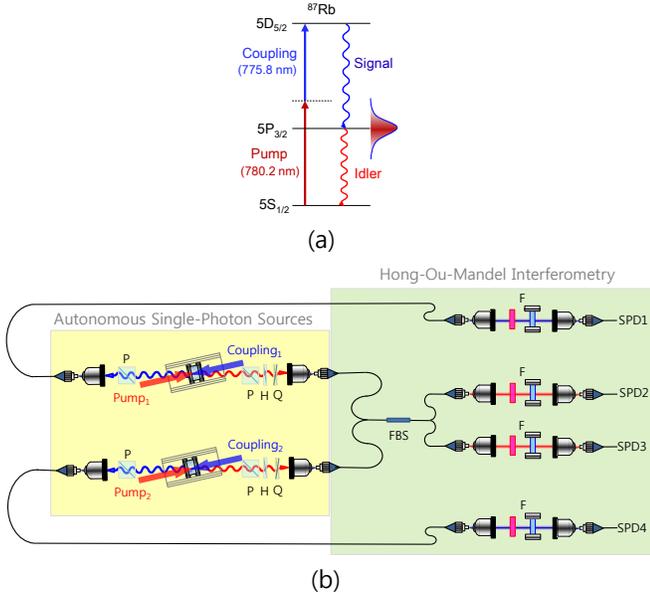

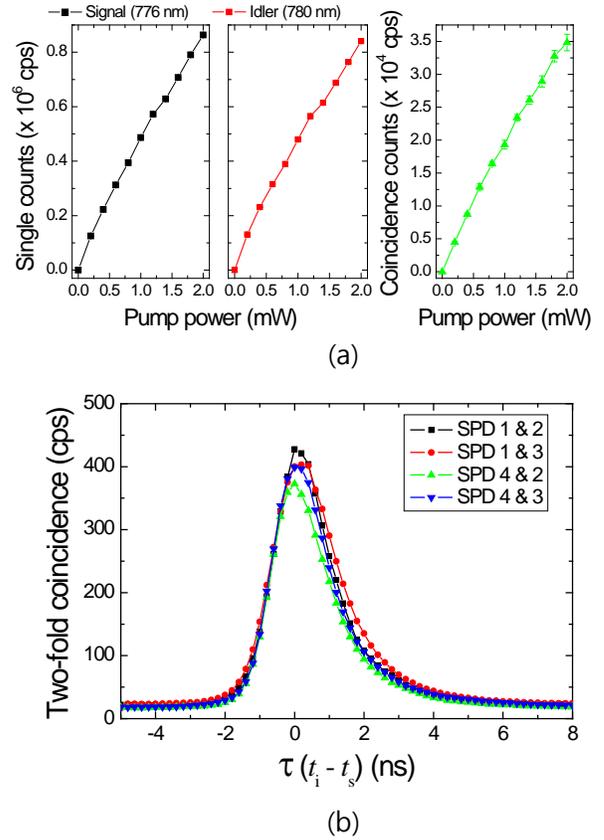

Fig. 1. (a) Generation of signal and idler photons owing to Doppler-free two-photon coherence via interaction of Doppler-broadened ladder-type atomic system with pump and coupling lasers. (b) Schematic of experimental setup for HOM interference between two independent CW photon pairs (P: polarizer; H: half-wave plate; Q: quarter-wave plate; FBS: fiber beam splitter; F: filtering stages composed of interference filter and solid fused-silica etalon filter; SPDs: single-photon detectors).

The experimental setup used to obtain HOM interference between autonomous photon pairs emitted from Doppler-broadened atomic ensembles is shown in Fig. 1(b). To achieve HOM interference between two independent photon pairs, we prepared twin CW photon-pair sources in two independent Doppler-broadened atomic systems, where the $Signal_{1(2)}$ and $Idler_{1(2)}$ photons were generated by the horizontally polarized $Pump_{1(2)}$ and vertically polarized $Coupling_{1(2)}$ lasers, as indicated in the figure. The pump and coupling beams with the diameter of 1.5-mm were counter-propagated in an atomic vapor cell with 12.5-mm-long and 25-mm-diameter. The temperatures of the two vapor cells were stabilized to near 60°C. For each atomic ensemble, the signal and idler photons, which were propagated at a 1.3° tilting angle from the pump and coupling beam path axes, respectively, were coupled into single mode fibers (SMFs). The photon pairs were collectively enhanced in the phase-matched direction. The photons were detected by a single photon counting module (PerkinElmer SPCM-AQRH-13HC).

In our experiment, the bright CW photon-pair sources were used to increase the rates of the four-fold events between two independent photon pairs. Figure 2(a) shows the signal (black squares) and idler (red squares) single counting rates and the net coincidence counting rate (green triangles) as functions of the pump power for a 10-mW coupling power. Note that detector dead times of approximately 50 ns were considered with no other corrections, e.g., optical loss, filter transmittance, fiber coupling, or detection efficiency adjustments. For 1-mW pump power and with the coupling power set to 10 mW, the net coincidence counting rate was measured to be approximately $2 \times 10^4$

Fig. 2. (a) Signal- and idler-photon single counting rates and photon-pair coincidence counting rate as functions of pump power for 10-mW coupling power. (b) Two-fold coincidence measurements between signal and idler photons yielded by four SPDs positioned after FBS; SPD1 and SPD4 detected $Signal_1$ and $Signal_2$, respectively, and SPD2 and SPD3 simultaneously detected $Idler_1$ and $Idler_2$ split and mixed by the FBS.

For the HOM interferometry setup shown in Fig. 1(b), the two idler photons were coupled with two SMFs and collected by a fiber beam splitter (FBS), in order to achieve spatial overlap between the independent single photons emitted from the sources. To match the polarization between these independent single photons, the polarizations of the idler photons were adjusted by half- and quarter-wave plates positioned before two input collimators.

In particular, the spectral features of our autonomous CW photon-pair source are intrinsically determined by the Doppler broadening due to the velocity distribution of the atomic ensemble. When the temperatures of both vapor cells are identical, almost full spectral overlap between the two independent single photon sources is obtained, because the photon-pair spectrum is determined by the Doppler effect. Therefore, very high spectral purity is obtained for the photon sources under the same vapor-cell temperature conditions and without supplemental filters. In order to remove the pump and coupling lasers, along with the uncorrelated fluorescence, we used interference and etalon filters. The photon-pair spectral width was estimated to correspond to approximately 100 MHz though wavelength mismatching and Doppler-broadening effects [22]. The bandwidth of the interference filters was 3 nm and the etalon filters had a full width at half maximum (FWHM) linewidth of 950 MHz. The spectral widths of photon pairs were smaller than those of the filters. There was no spectral narrowing of the photon pairs due to the filters. After passing through the filters, the signal and idler photons were sent to four SPDs, which were connected to a time tagging module for four-fold coincidence measurement.

Temporal mode matching of the independent photon pairs is important for obtaining the high visibility HOM interference. Figure 2(b) shows the two-fold coincidence counts between the four SPDs located after the FBS, where $\tau$ is the relative delay of the idler photons (measured by SPD2 and SPD3) relative to the two independent signal photons (measured by SPD1 and SPD4). From the results of Fig. 2(b), we determined the characteristics of the four SPDs and the temporal mode between the signal and idler photons emitted from the two separate atomic vapor cells. However, the peak values of the coincidence counts differed slightly, because of the different detection efficiencies of the four SPDs. The temporal shapes correspond to the cross-correlation function between the signal and idler photons, which is determined by measuring the coincident detection histogram as a function of $\tau$. The FWHM widths of the temporal shapes were estimated to be approximately 2 ns. The photon-pair coherence time corresponded to approximately 4 ns; this is defined as the relative time for which the coincidence counts decayed as $1/e^2$. To check the SPD time resolution, we measured the time resolutions of the four SPDs using a weak mode-locked picosecond-pulse laser. The time uncertainties of the SPDs were measured from 0.42 to 0.49 ns, including the electronic timing jitter. Note that the photon-pair temporal coherence time is so short that the trigger photon of the heralded single photon is measured with a large time uncertainty. However, the purity of the single-photon source may be limited by the time uncertainties of the SPDs, because the purity of the single-photon source is determined by a product of temporal biphoton bandwidth and the resolution between SPDs.

Figure 3 shows the HOM interference obtained using two independent single-photon pairs in a 900-s effective measurement time with a 7-ns temporal window. The x-axis indicates the arrival time difference ($\Delta t$) for the two independent single-photon pairs, where the time step of $\Delta t$ is set to 0.4 ns. The HOM interference visibility is estimated to be 83%. Through use of the bright autonomous CW-mode photon pairs, we obtained higher-rate four-fold photon coincidence events, corresponding to 200 events/h for a 0.4-ns temporal range.

When two independent photon pairs are identical in terms of their spatial and polarization modes at the FBS, the HOM interference between these two independent single-photon pairs is described as

$$P^{(4)}(\Delta t) = T^2 \int d\tau |h_1(\tau)|^2 \int d\tau |h_2(\tau)|^2 + R^2 \int d\tau |h_1(\tau)|^2 \int d\tau |h_2(\tau)|^2 \\ -2TR \left| \int d\tau h_1^*(\tau) h_2(\tau + \Delta t) \right|^2 \\ +4TR\tilde{g}^{(2)} \frac{\int d\tau |h_1(\tau)|^2}{2} \frac{\int d\tau |h_2(\tau)|^2}{2}, \quad (1)$$

where $h_{1(2)}(\tau) = \langle 0 | \hat{a}_{i1(2)}(t+\tau) \hat{a}_{s1(2)}(t) | \Psi \rangle$ is the temporal wave function amplitude and $\tilde{g}^{(2)}$ is the averaged heralded auto-correlation within the coherence time window [3]. The red curve of Fig. 3 represents the theoretical HOM curve, for which $\tilde{g}^{(2)}$ was determined to be 0.356 through measurement of the heralded auto-correlation for a 7-ns temporal window. The experimental and theoretical results are in good agreement. The purity of the heralded single photons from the independent warm atomic ensemble sources was 0.973±0.029. This purity is in good agreement with the calculated purity value of 0.986 [23]. See the Supplementary Material for further details [24].

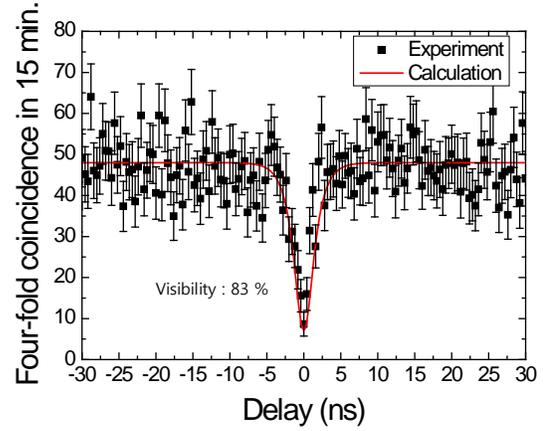

Fig. 3. Four-fold coincidence measurement of two independent CW photon pairs in 900-s measurement period. The HOM interference has 83 % visibility.

In conclusion, we observed the HOM interference between two autonomous CW-mode photon pair sources generated from a Doppler-broadened ladder-type $^{87}$Rb atomic system. Highly bright photon-pair sources were obtained from the Doppler-free two-photon coherence via interaction of Doppler-broadened atomic ensembles with coherent electromagnetic fields of similar wavelength. Furthermore, very high spectral purity was achieved for our photon-pair source in the absence of filters, because the spectral features of the photons correspond to a Gaussian profile, which results from the Doppler effect in the warm atomic ensemble. In a short measurement period of 900 s, we successfully demonstrated HOM interference with visibility as high as 80%, using bright autonomous photon-pair sources in a warm atomic ensemble for the first time. We believe that our results can contribute to the realization of long-distance quantum communication and optical quantum-information processing, using linear gates or through the formation of cluster states.

**Acknowledgment.** We are grateful to J. F. Chen and S. M. Lee for fruitful discussions.